# Analysis of Degradation in Graphene-based Spin Valves

Kazuya Muramoto [1,**], Masashi Shiraishi [1,2,*,**], Nobuhiko Mitoma [1], Takayuki Nozaki [1], Teruya Shinjo [1], Yoshishige Suzuki [1]

1. Graduate School of Engineering Science, Osaka Univ.
   1-3 Machikaneyama-cho, Toyonaka 560-8531, Osaka, Japan
2. PRESTO-JST, 4-1-8 Honcho, Kawaguchi 332-0012, Saitama, Japan.

* Corresponding author) Masashi Shiraishi
   E-mail) shiraishi@mp.es.osaka-u.ac.jp
** These two authors contributed equally to this work.

## Abstract

The degradation mechanisms of multilayer graphene spin valves are investigated. The spin injection signals in graphene spin valves have been reported to be linearly dependent on the drain bias voltage, which indicates that the spin polarization of injected spins in graphene is robust against the bias voltage. We present that the disappearance of this robustness is due to two different degradation mechanisms of the spin valves. Our findings indicate that the disappearance of the robustness is due to degradation rather than an intrinsic characteristic of graphene. Thus, the robustness can be greatly enhanced if degradation can be prevented.

Graphene is a critical material in the field of molecular spintronics because reproducible spin injection and generation of a pure spin current at room temperature have been demonstrated [1–3]. After the first success of the spin injection into single- and multi-layer graphene (SLG and MLG) at RT by several groups [1-3], a number of important findings have been reported, such as anisotropic spin relaxation [4], a spin drift motion [5], robustness of spin polarization of injected spins [6], comparably long spin coherence [2,6,7] and spin manipulation (the Hanle effect) [2,6,8]. These findings represent essential building blocks for further progress in molecular spintronics based on graphene, whereas they has not been achieved in other molecular spin devices using carbon nanotubes, Alq3 and so on.

Our group has observed a pure spin current and robust spin polarization of injected spins in MLG spin valves [1,6], and also observed gate-induced modulation of spin signals in single-layer graphene spin valves with transparent contacts, as a theory teaches us [9]. The robustness of the spin polarization in local and non-local schemes is a characteristic of spin transport in graphene that has not been observed in other metallic and semiconductor systems besides Si [10]. The robustness is exhibited in the linear dependence of the spin injection signal on the drain bias voltage (i.e., on the injection electric current). The spin injection signal is defined as $\Delta V_{non-local} = (V_P - V_{AP})$, where $V_P$ and $V_{AP}$ are the output voltages for parallel and antiparallel magnetization alignments, respectively. The robustness of the spin polarization is considered to be due to the spin polarization of spins injected into graphene remaining constant. This is in accordance with the following equation for a non-local spin signal described by Takahashi and Maekawa [11]:

$$\Delta V_{non-local} = \frac{2P^2}{(1-P^2)^2}(\frac{R_F}{R_N})R_F \cdot [\sinh(\frac{L}{\lambda_{sf}})]^{-1} \cdot I_{inject} \quad , \qquad (1)$$

where $\Delta V_{non-local}$ is the spin signal, $P$ is the spin polarization, $\lambda_{sf}$ is the spin flip length, $L$ is the gap length between the two ferromagnetic (FM) electrodes, $I_{inject}$ is the injected electric current, and $R_F$ and $R_N$ are respectively the spin accumulation resistances of a ferromagnet and a nonmagnet, which are defined as (conductivity)×(spin diffusion length)/(cross-sectional area). The observed robustness is a very significant finding for graphene spintronics, especially for graphene spin transistors, because it implies that the magnetoresistance ratio will not be reduced when a bias voltage is applied and that a possibility of device designing can be greatly expanded.

However, further investigations have revealed that graphene-based spin devices sometimes exhibit a sublinear dependence of the spin signals on the high drain bias voltage (i.e., at high injection electric currents). It was pointed that this sublinearity would be due to degradation of the interface between the FM electrodes and the graphene [6], but the detailed characteristics of this mechanism have not been clarified. In this study, we performed a detailed analysis of the sublinear dependence of the spin signals. We found that there are two types of degradation at the interface between the FM electrodes and the graphene and that the sublinear dependence is not an intrinsic physical characteristic of graphene.

The starting materials used for preparation of the MLG spin valves were highly oriented pyrolytic graphite (HOPG, NT-MDT Co.) and polyimide-oriented highly oriented graphite (Super graphite, Kaneka Co.) [12]. Adhesive tape was used to peel MLG flakes from these materials and the flakes were then pressed onto the surface of a SiO₂/Si substrate (SiO₂ thickness = 300 nm). The typical thickness of MLG flakes was 2–40 nm. The non-magnetic and FM electrodes used were Au/Cr (40 nm/5 nm) and Co (50 nm), respectively, and they were patterned by electron-beam lithography using a ZEP-520A resist. The two Co electrodes (Co1 and Co2) had the same widths, but the Co2 electrode had a pad-like structure in order to weaken the coercive force (see Fig. 1(a)). The gap between these two Co electrodes was typically 1.5 μm. All magnetoresistance measurements were performed at room temperature.

We introduced a non-local scheme [13] where one can detect a non-local output voltage which is induced by position dependence of electrochemical potential of the generated spin current in the MLG (see Fig. 1(b)). The spin injection was investigated using a four-probe system (ST-500, Janis Research Co. Inc.) with an electromagnet. A source meter (KH2400, Keithley Instruments Inc.) and a multimeter (KH2010, Keithley Instruments Inc.) were used to detect the spin injection signals. Here, spins are injected as an electric current at the ferromagnet/graphene interface on the top layer of the MLG on the injector side (between the Co1 and Au1 electrodes, see Fig. 1(a)). Accumulated spins diffuse from the Co1 electrode to the Co2 and Au2 electrodes (on the detector side). Hence, the output voltage that is induced by the generated spin current is determined by the amount of accumulated spins; namely, it is strongly affected by the spin polarization at the Co1/graphene interface. In a non-local scheme, an electric current $I$ is injected from the Co1 electrode into MLG and extracted at the Au1 electrode; this is termed the positive bias condition. The non-local output voltage is measured between the Co2 and Au2 electrodes. The intensity of the non-local spin signal, $\Delta V_{non\text{-}local}$, is defined as the difference between the spin signal intensities in the parallel and antiparallel magnetization alignments of the two FM electrodes (i.e., $\Delta V_{non\text{-}local} = V_P - V_{AP}$) as shown in Fig. 1(b).

Figure 2(a) shows an example of the sublinearity of the spin signal in a MLG spin valve. It shows that the spin signal loses its linearity at about +8 mA (positively biased) where its intensity decreases dramatically. Figure 2(b) shows the dependence of the MLG resistance between the Co1 and Co2 electrodes measured by the four-probe method, and Fig. 2(c) shows the dependence of the Co1−MLG−Co2 resistance measured by the two-probe method. The following procedure was used to measure the spin injection and the sample resistance:

Step 1: spin signal detection at 1 mA

Step 2: four-probe measurement of the MLG resistance

Step 3: two-probe measurement of the Co1−MLG−Co2 circuit resistance

Step 4: spin signal detection at 2 mA

Step 5: four-probe measurement of the MLG resistance

………

Step 34: spin signal detection at 12 mA

Step 35: four-probe measurement of the MLG resistance

Step 36: two-probe measurement of the Co1−MLG−Co2 circuit resistance

Note that the horizontal axis in Figs. 2(b) and (c) indicates the electric current for spin injection into the Co1−MLG−Au1 circuit. The Co1−MLG−Co2 circuit resistance is defined as: $R_{Co1−MLG−Co2}$ = (detected bias voltage)/(1 mA). The Co1/MLG interface is the only component where the electric current for both spin injection and the measurement of $R_{Co1−MLG−Co2}$ flows in the measurement sequence. The MLG resistance remained constant, whereas the resistance of the Co1−MLG−Co2 circuit measured by the two-probe method increased suddenly at a spin injection current of 7 mA. The only plausible explanation for this steep increase is an increase in the interface resistance between the Co1 electrode and the MLG. The increase in the interface resistance is probably due to degradation caused by Joule heating and the disappearance of a distinct interface due to the application of a high bias voltage, which can induce spin scattering.

However, this mechanism was not the only degradation mechanism observed in the MLG spin valves; another mechanism was observed, as shown in Figs. 3(a)−(c). Here, there are two interesting features: (1) no dramatic reduction in the spin signal was observed, whereas sublinearity occurred at an injected electric current of above 12 mA, and (2) the Co1−MLG−Co2 resistance measured by the two-probe method decreased rather than increased above 12 mA. The second feature is the opposite tendency to that for the degradation mechanism discussed above. To investigate this mechanism, we prepared another spin valve and measured the $I$−$V$ curves of the Au1−MLG−Co1 circuit under various conditions. Figure 4(a) shows the dependence of the spin signal on the bias voltage and Fig. 4(b) shows the $I$−$V$ curves of the Au1−MLG−Co1 circuit after electric currents of 1 mA and 4 mA were injected for observing spin signals. The $I$−$V$ curve did not exhibit ohmic characteristics for the lower injection current (1 mA, shown by the red curve), whereas ohmic characteristics were observed at the higher injection current (5 mA, shown by the blue curve). The non-ohmic character most likely indicates the existence of a tunneling barrier between the Co1 electrode and graphene, which was probably formed by residual ZEP-520A resist from the fabrication process. The disappearance of the non-ohmic characteristics is due to this barrier being broken. According to theory [11], such a relationship implies that there is a serious conductance mismatch in this system and that a reduction in the tunnel resistance induced the reduction in the spin signals due to the difficulty in achieving efficient spin injection.

Superficially, these two mechanisms appear contradictory. However, it should be noted that both mechanisms do not occur simultaneously and the second degradation mechanism was found to be the principal mechanism. The most important finding of these experiments is that much more robust spin polarization of injected spins can be obtained if such degradation can be suppressed. In other words, the disappearance of the robustness is not due to an intrinsic characteristic of graphene, but rather it is due to an extrinsic cause, namely degradation.

The authors thank Dr. T. Takano and Dr. T. Seki for fruitful discussions and Mr. T. Sugimura for his assistance with the experiments.

**Figure Captions**

**Figure 1**

(a) An optical microscopic image of a MLG spin valve. (b) A schematic view of the measurement circuit of spin injection into the MLG spin valves and a typical spin injection signal at an injection electric current of +1 mA. The electric current is injected from the Co1 electrode to the Au1 electrode (defined as positively biased in this study). $\Delta V_{non-local}$ is defined as the difference between the spin injection signal intensities for parallel and antiparallel magnetization alignments.

**Figure 2**

(a) Dependence of the spin signal on the injected electric current in the MLG spin valve. (b) The MLG resistance as a function of the injection electric current for spin injection. The resistance was measured by the four-probe method and remained constant in these measurements. (c) The resistance of the Co1–MLG–Co2 circuit as a function of the injection electric current for spin injection. The resistance was measured by the two-probe method and it increased dramatically at about 8 mA. The horizontal axes in (b) and (c) indicate the electric current for spin injection. The red line in all of figures is a visual guide.

**Figure 3**

(a) Dependence of the injection electric current on the spin signals in the MLG spin valve. (b) The MLG resistance as a function of the injection electric current for spin injection. The resistance was measured by the four-probe method and remained constant in these measurements. (c) The resistance of the Co1−MLG−Co2 circuit as a function of the injection electric current for spin injection. The resistance was measured by the two-probe method and decreased gradually after about 12 mA. The horizontal axis in (b) and (c) indicates the electric current for spin injection. The red line in all of the figures is a visual guide.

**Figure 4**

(a) Dependence of the spin signal on the injection electric current. The colors of the closed circles correspond to the colors of the $I$−$V$ curves in (b). (b) $I$−$V$ curves of the Co1−MLG−Au1 circuit (i.e., the electric current injection circuit) in the MLG spin valve. The red and blue curves are the $I$−$V$ curves after 1 mA (before degradation) and 5 mA (after degradation) were applied for spin injection, respectively. Non-ohmic behavior is observed prior to degradation, whereas ohmic behavior appears after degradation.

**Figures (*Color online*)**

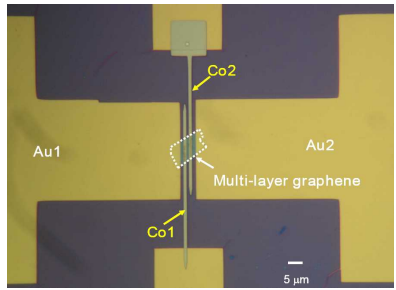

**Fig. 1(a)**

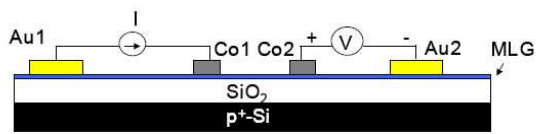

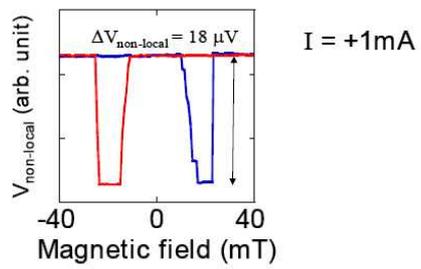

**Fig. 1(b)**

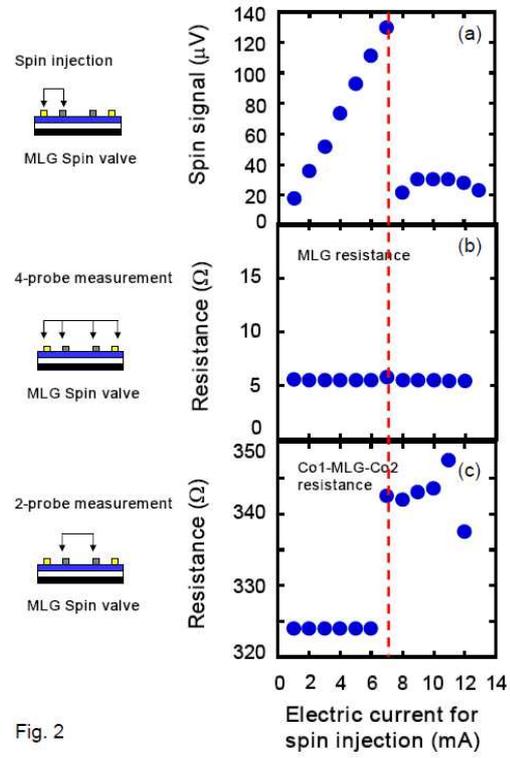

Spin injection

MLG Spin valve

4-probe measurement

MLG Spin valve

2-probe measurement

MLG Spin valve

Fig. 2

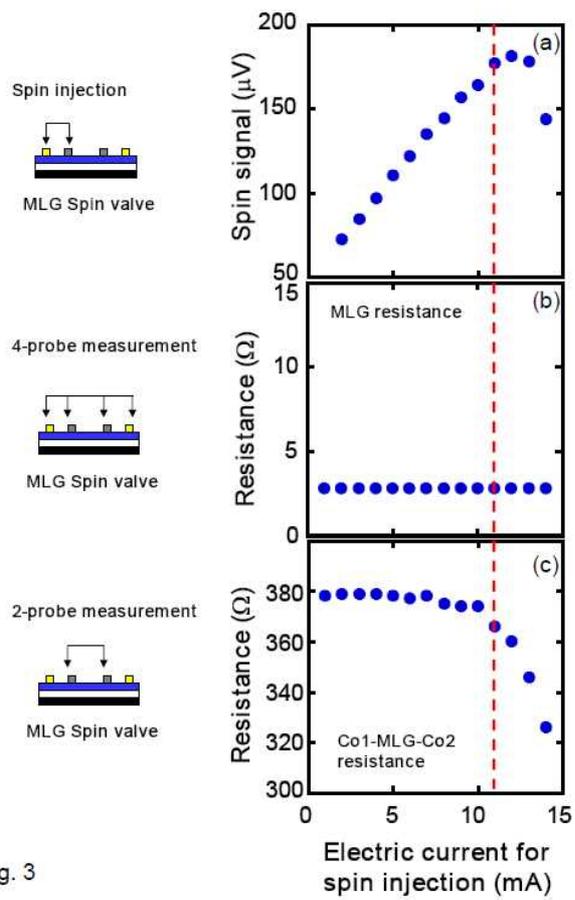

Spin injection

MLG Spin valve

4-probe measurement

MLG Spin valve

2-probe measurement

MLG Spin valve

Fig. 3

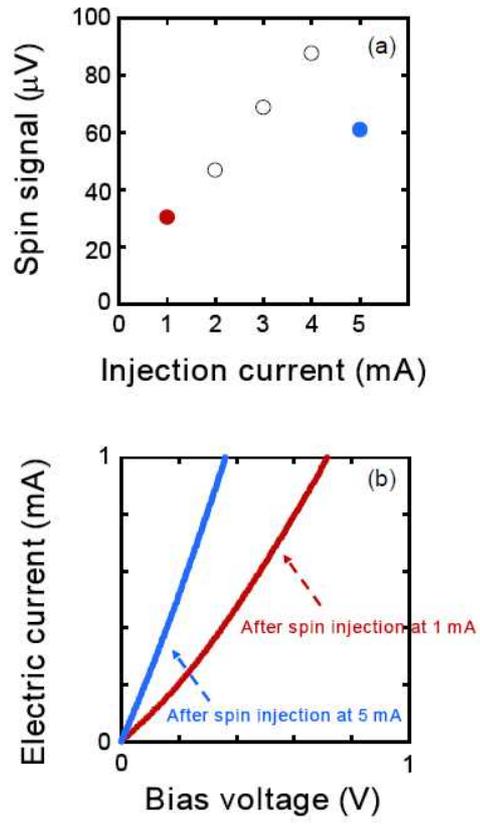

Fig. 4